\documentstyle[12pt,aasms4]{article}

\lefthead{Fuente A., Mart\'{\i}n-Pintado J.}
\righthead{Detection of CO$^+$...}

\begin{document}

\title{Detection of CO$^+$ toward the reflection nebula NGC 7023
}

\author{A. Fuente and J. Mart\'{\i}n-Pintado}
\affil{Observatorio Astron\'omico Nacional (IGN), Campus Universitario,
Apdo. 1143, E-28800 Alcal\'a de Henares (Madrid), Spain }

\begin{abstract}
We have detected CO$^+$ toward the photon-dominated region (PDR)
associated with the reflection nebula NGC 7023. 
This is the first detection of CO$^+$ in the vicinity of a Be star.
A CO$^+$ column density of $\sim$ 3 10$^{11}$  cm$^{-2}$ has been 
derived toward the PDR peak.
We have, however, not detected CO$^+$ in a well shielded clump of 
the adjacent molecular cloud, where the CO$^+$/HCO$^+$ abundance
ratio is {\it at least 100 times lower} than in the PDR.
Our results show, for the first time, that CO$^+$ column densities
as large as $\sim$ 3 10$^{11}$  cm$^{-2}$ can be produced 
in regions with incident UV fields of just a few 10$^3$ (in units
of Habing field)
and densities of $\leq$ 10$^5$ cm$^{-3}$. Furthermore, since the
ionization potential of CO is larger than 13.6 eV, our data rule out the 
direct photoionization of CO as a significant CO$^+$ formation mechanism. 

\end{abstract}


\section{Introduction}
Although CO is the most abundant interstellar molecule after H$_2$, its 
corresponding ion, CO$^+$, is expected to have very low abundance
in molecular clouds. The reason is that CO$^+$
is quickly converted into HCO$^+$ by reactions with H$_2$.
Only in the hot layers of photon-dominated regions (PDRs) where
a significant fraction of hydrogen is still in atomic form, the
CO$^+$ abundance becomes significant. Chemical models
(\cite{ste95}) predict that for the PDRs associated with massive stars
(n $\sim$ 10$^6$ cm$^{-3}$, G$_\circ$ $\sim$ 2 10$^5$ in units of Habing
field) the CO$^+$/HCO$^+$
abundance ratio is $\sim$0.05 at a visual extinction lower than 1.5 mag,
but decreases by more than 2 orders of magnitude when the extinction
increases above 3-5 mag. Based on its chemical behavior,
they proposed the CO$^+$/HCO$^+$ ratio as a tracer of
the HI/H$_2$ transition layer in PDRs. 

CO$^+$ was tentatively detected for the first time by \cite{eri81}
toward OMC-1. Later, \cite{lat93} detected CO$^+$ in the well-known
PDR M17SW and the planetary nebula NGC 7027. More recently, CO$^+$ has 
also been detected in the Orion Bar (\cite{sto95}).
But so far, all the detections of CO$^+$ have
been made toward the interfaces between the molecular cloud
and the HII regions around massive O stars. \cite{sto95} failed to
detect CO$^+$ toward the reflection nebula S140. They propose that 
the large densities and intense UV fields associated with massive
O stars are required to form CO$^+$ column densities 
$\ge$ 10$^{11}$ cm$^{-2}$.
We present the detection of CO$^+$ toward the reflection nebula
NGC 7023, which is illuminated by a Be star.
Although in this region the incident UV field is, 
$G_\circ$ $\sim$ 10$^3$ (in units of Habing field),  
and densities are, n $\sim$ 10$^5$ cm$^{-3}$, we have observed a 
CO$^+$ column density of $\sim$ 3 10$^{11}$ cm$^{-2}$.
Furthermore, the spatial-velocity
distribution of the CO$^+$ emission shows that CO$^+$ is only 
located within the HI/H$_2$ transition layer of this PDR.

\section{Observations and Results}

In Plate L1, we show the integrated intensity map of 
the HCO$^+$ 1$\rightarrow$0 
(thin dark contours) and the HI column density map (grey scale)
of the reflection nebula NGC 7023 (\cite{fue93} and \cite{fue96}). 
The illuminating star, HD 200775,
is located in a cavity of the parent cloud
delimited by dense walls (\cite{fue93} and references therein).
Dense PDRs are found on the surfaces of these walls.
In particular, an intense HI clump appears
$\approx$ 40 NW from the star position, at the edge of the bulk of
the molecular emission (see Plate L1). 
Interferometric observations of the J= 1$\rightarrow$0 line of HCO$^+$
showed the existence of several high 
density HCO$^+$ filaments within this clump (\cite{fue96}). 
The two most intense filaments are also shown in Plate L1.          
We have searched for CO$^+$ toward the peak position of these 
filaments. The coordinates of this position are given in Table 1 and
hereafter, we will refer to it as the ``PDR peak".
Beyond the ``PDR peak", our HCO$^+$ single-dish data show the existence 
of several clumps immersed in the molecular cloud. 
We have also searched for CO$^+$
toward the molecular clump closest to the ``PDR peak" (Plate L1). 
The coordinates of this position are also given in Table 1 and
hereafter, we will refer to it as the ``Molecular Peak''. 
 
CO$^+$ has a $^2$$\Sigma$
ground electronic state in which each rotational level is split in
two fine structure levels with J=N$\pm$1/2. The N=1$\rightarrow$0 
rotational line
is heavily obscured by the O$_2$ line at 118 GHz and cannot be observed
from ground-based telescopes. The most intense transitions of the 
N=2$\rightarrow$1 rotational spectrum are N=2$\rightarrow$1 
J=5/2$\rightarrow$3/2 at 236062.553 MHz and N=2$\rightarrow$1 
J=3/2$\rightarrow$1/2 at 235789.641 MHz. In the optically thin limit,
the intensity ratio I(235.789)/I(236.062) is 0.55 (\cite{sas81}).
Both line frequencies were covered by the receiver band.
Unfortunate, the most intense line is blended with 
the 5$_{-2}$$\rightarrow$4$_{-2}$ and  5$_{2}$$\rightarrow$4$_{2}$ 
$^{13}$CH$_3$OH E lines (see \cite{ble84}). In order to determine an
upper limit to the $^{13}$CH$_3$OH emission we have observed 
the 5$_1$$\rightarrow$4$_1$ methanol line toward
the PDR peak. To determine accurate CO$^+$ column densities,
it is necessary to have accurate estimates of the hydrogen density. For this 
aim, maps of about 20''$\times$20'' with a spacing
of 5'' were carried out around the studied positions 
in the CS J=2$\rightarrow$1, 
3$\rightarrow$2 and 5$\rightarrow$4 lines. Furthermore, the H$^{13}$CO$^+$
J=1$\rightarrow$0 (toward both positions)  and 3$\rightarrow$2 (only toward
the molecular peak) have also been observed. 

The observations were carried out in 1995 December and 1996 May
using the 30-m telescope. The observational procedure was position
switching with a fixed reference 30' East from the star.
Pointing was checked every two hours using strong continuum sources
(NGC 7027, K3-50A, NGC 7538), and the rms of pointing errors was less 
than 2''. The forward and main beam efficiencies were 0.92 and 0.75
at 90 GHz, 0.90 and 0.52 at 145 GHz, and 0.86 and 0.37 at 236-260 GHz
respectively.
The temperature scale is main beam temperature. The HPBW of the
telescope was 27'' at 90 GHz, 16'' at 145 GHz and 10'' at 236 GHz.  
Typical system temperatures (in T$_{MB}$) were 300 K at 90 GHz, 
600 K at 145 GHz, 1300K at 236 GHz and 3400 at 260 GHz. All the lines
have been observed with a frequency resolution of 80 kHz ($\sim$ 0.1
km s$^{-1}$ at 236 GHz). 

\placetable{tbl-1}

\subsection{PDR peak}
     
The CO$^+$ N=2$\rightarrow$1 5/2$\rightarrow$3/2 and 3/2$\rightarrow$1/2 lines
have been detected toward the PDR peak with a signal to noise ratio
of 10 and 7 respectively (see Plate L1). The observational parameters are
shown in Table 1. The detection of both lines
make very unlikely a possible misidentification. We are not aware
of possible line contamination for the transition at 235789.64 MHz. 
The only possible line contamination comes from
the 5$_{-2}$$\rightarrow$4$_{-2}$ and  5$_{2}$$\rightarrow$4$_{2}$ 
$^{13}$CH$_3$OH E lines whose rest frequency is less than 0.5 MHz from 
the CO$^+$ line at 236062.55 MHz (see Blake et al. 1984). Since
the observed linewidths of the two lines of CO$^+$ are the same, it 
seems that the line at 236062.55 MHz is very unlikely contaminated by
$^{13}$CH$_3$OH lines. To check for possible contamination,
we have estimated an upper limit to the 
emission of the $^{13}$CH$_3$OH lines from
the observed J$_K$=5$_1$$\rightarrow$4$_1$ methanol line. The 
excitation conditions and the line strength for this line 
are very similar to those of the contaminating $^{13}$CH$_3$OH lines
(\cite{and90}). 
Assuming a linewidth of 2 kms$^{-1}$,
we have obtained a 3$\sigma$ upper limit of 0.2 K kms$^{-1}$ to the 
integrated intensity emission of the methanol line (see Table 1). 
For an isotopic ratio, CH$_3$OH/$^{13}$CH$_3$OH $\sim$ 40, 
this would imply an upper limit of 0.005 K kms$^{-1}$ to the integrated 
intensity emission of each $^{13}$CH$_3$OH line. Since
there are two $^{13}$CH$_3$OH lines blended, 
these lines could contribute to our CO$^+$ detection
at 236.062 GHz with, at most, an integrated intensity of 0.01 K km s$^{-1}$.
This is only 4\% of the observed integrated intensity emission
at 236.062 GHz, and it is 
within the observational errors (see Table 1). 
We, therefore, conclude
that the emission detected at 236.062 and 235.789 GHz is 
due to CO$^+$.

A striking result of our data is that the CO$^+$ lines have 
linewidths much larger
than those of CS and H$^{13}$CO$^+$. The interferometric
HCO$^+$ filaments detected toward the PDR peak
are characterized by having
different velocities.
The four well detected filaments are centered
at radial velocities of 1.9, 2.4, 2.8 and 4 km s$^{-1}$,
and the one tentatively detected is centered at 5.8  kms$^{-1}$.
One of these filaments, 2.4 kms$^{-1}$, is
very likely embedded in the molecular cloud, but most of the
others, 2.8, 4.0 and 5.8 km s$^{-1}$, seem to be immersed in the 
atomic medium. The situation
is less clear for the filament at 1.9 kms$^{-1}$ that seems to be
part of a weak and extended molecular component (\cite{fue96}). Therefore,
a gradient in the chemical composition of the HCO$^+$ filaments
is expected depending upon the local visual extinction toward the exciting
star.
CS and H$^{13}$CO$^+$ present
narrow lines centered at 2.4 kms$^{-1}$, i.e., the velocity of the
filament immersed in the bulk of the molecular cloud. 
Only HCO$^+$ and CO$^+$ present emission
at the velocities of the filaments immersed in the atomic medium. 
From the comparison of the
spectra of the H$^{13}$CO$^+$, HCO$^+$ and CO$^+$ lines,
it is clear that there exists a gradient 
in the CO$^+$/HCO$^+$ abundance 
ratio as a function of velocity, i.e., as a function of 
the visual extinction from the star
(see Plate L1). To determine this gradient,
we have estimated the CO$^+$/HCO$^+$ abundance 
ratio in 
three different velocity intervals, 0 - 1.6 km s$^{-1}$, 
1.6 - 3.2 km s$^{-1}$, and 3.2 - 6 km s$^{-1}$.
    
CO$^+$ column densities have been estimated 
using the LTE approximation.
Assuming T$_K$ = 40 K (see \cite{fue93}) we derived from CS data
a hydrogen density of $\sim$3.5 10$^5$ cm$^{-3}$ for the component at 
2.4 km s$^{-1}$. Similar densities were obtained for 
the other filaments
from the interferometric HCO$^+$ data (\cite{fue96}).
Using a LVG code and assuming T$_k$ = 40 K and 
n = 3.5 10$^5$ cm$^{-3}$, we estimate T$_{rot}$ = 10 K for HCO$^+$.
Since HCO$^+$ and CO$^+$ have similar dipole moments and rotational
constants ($\mu$ = 2.77 D for CO$^+$ and 3.91 D for HCO$^+$), 
we assume the same rotational temperature for CO$^+$.
In Table 2 we show the derived HCO$^+$, H$^{13}$CO$^+$ and CO$^+$
column densities. From these estimates, we have determined that
the CO$^+$/HCO$^+$ abundance 
ratio is a factor of 10 larger for the filaments 
immersed in the atomic
region than for the filaments embedded in the molecular cloud. 
This gradient in the CO$^+$/HCO$^+$ ratio cannot be due 
to an opacity effect.
The I(CO$^+$ 235.789)/I(CO$^+$ 236.062) ratio is consistent with
optically thin emission (within the observational errors) for all
the velocity intervals (see Table 2). Though consistent with 
optically thin emission, our data suggest that the opacities of
the CO$^+$ lines could be larger for the velocities 
3.2 - 6.0  km s$^{-1}$ than for 1.6 - 3.2  km s$^{-1}$. In this
case, the CO$^+$ column density would be slightly underestimated
for the velocity interval 3.2 - 6.0  km s$^{-1}$, and the
derived CO$^+$/HCO$^+$ ratio would be a lower limit to the actual
value of the CO$^+$/HCO$^+$ ratio for this interval. 
Therefore, although we are
aware of the uncertainties involved in column density estimates,
we think that the observed gradient in the CO$^+$/HCO$^+$ ratio 
(a factor of 10) is significant, and it is in agreement
with the expected behavior of the CO$^+$/HCO$^+$ ratio,
where CO$^+$ formation is restricted to a narrow range
of visual extinctions A$_v$ $<$ 2 mag. The visual extinction at the surface
of the filament at 2.4 km s$^{-1}$ must be $>$1 mag to be immersed
in a mainly molecular medium, while for the filaments immersed
in a mainly atomic medium, the visual extinction must be $<$1 mag. 
Assuming a HCO$^+$ fractional  
abundance of 4 10$^{-10}$ (\cite{fue96}), 
the CO$^+$ fractional abundance is 
$\sim$ 4 10$^{-11}$ in the filaments
immersed in the atomic medium.
CO$^+$ fractional abundances $\sim$ 10$^{-11}$ are also
derived from the CO$^+$ data reported by \cite{sto95}
and \cite{lat93}, toward M17SW and the Orion Optical Bar.
Although the physical conditions and incident UV field 
are different (see Section 3), the CO$^+$ fractional abundance 
in NGC 7023 is
similar to that found at the edges of the HII regions around
massive stars. 

\placetable{tbl-2}

\subsection{Molecular peak}
 
We have not detected CO$^+$ toward the molecular peak. 
Assuming a linewidth of 1 km s$^{-1}$ (a typical linewidth
for the molecular cloud),  
we obtain an upper limit to the
integrated intensity of the CO$^+$ line at 236.062 GHz 
of 0.03 K kms$^{-1}$. 
Assuming a kinetic temperature of T$_K$ = 15 K (\cite{fue90}), 
we estimate a density of 10$^5$ cm$^{-3}$ from our CS data. 
This density is high enough to excite the CO$^+$ lines. In fact,
the excitation conditions required for the 
H$^{13}$CO$^+$ J=3$\rightarrow$2 line are comparable 
to those required for 
the CO$^+$ N=2$\rightarrow$1 J=5/2$\rightarrow$3/2 
and 3/2$\rightarrow$1/2 lines, 
and the H$^{13}$CO$^+$ J=3$\rightarrow$2 line 
has been detected with an intensity of 1.04 K. 
Therefore, the lack of detection of CO$^+$ toward the
molecular peak is not due to the excitation conditions 
in this region. 
With the same assumptions as for the PDR peak, 
the upper limit to the CO$^+$ column density 
is 4.5 10$^{10}$ cm$^{-2}$.
Assuming n = 10$^5$ cm$^{-3}$ and T$_K$ = 15 K, 
we estimate a H$^{13}$CO$^+$ column density of
8 10$^{11}$ cm$^{-2}$. This means a CO$^+$/HCO$^+$ ratio of $<$ 0.001. 
Therefore, the 
CO$^+$/H$^{13}$CO$^+$ ratio is {\it at least 100 times lower} in 
the molecular peak than in the filaments immersed in the atomic medium.
Assuming a HCO$^+$ abundance of 4 10$^{-10}$, we obtain a fractional
abundance of CO$^+$ of $<$5 10$^{-13}$ in the molecular peak. 
  
\section{Summary and Discussion}

We have detected, for the first time, CO$^+$ in a PDR associated with
a Be star. This region is very different from the massive star forming 
regions where CO$^+$ had been detected thus far. First of all, since
the ionization potential of CO is larger than 13.6 eV, a Be star does 
not produce a significant number of photons capable to ionize CO. 
Furthermore, the intensity of the UV field and the density
around this star, $G_\circ$$\sim$ 10$^3$ ( in units of Habing field), and 
densities of $\sim$ 10$^5$ cm$^{-3}$,
are very different from those around massive O stars where
$G_\circ$$\sim$ 10$^5$ and n$\ge$ 10$^6$ cm$^{-3}$. 
Chemical models predict that CO$^+$ column densities
decrease sharply for UV fields $<$10$^5$, and
densities $<$10$^6$ cm$^{-3}$ (\cite{sto95}). 
Even for the conditions prevailing in massive star forming regions, 
chemical models fail to predict the
large CO$^+$ column densities observed toward them. 
To solve this problem, some authors have suggested that
the direct photoionization
of CO might be a non-negligible formation mechanism of CO$^+$ in
these regions (\cite{jan95}, \cite{bla96}). 
We have estimated a CO$^+$ column density of $\sim$ 3 10$^{11}$ 
cm$^{-2}$ toward the PDR peak in NGC 7023.
Our results show that large CO$^+$ column densities can be produced
even with UV fields of just a few 10$^3$ and densities of around
10$^5$ cm$^{-3}$. Since the peak CO$^+$ abundance in NGC 7023 
($\sim$ 4 10$^{-11}$) is similar
to that found in massive star forming regions, our data suggest 
that the direct photoionization of
CO is not a significant formation mechanism for CO$^+$.  


\acknowledgments

We are grateful to the technical staff of Pico de Veleta for their
support during the observations. We are also grateful to Dr. R. Gaume
for his careful reading of the manuscript.
This work has been partially supported by the Spanish 
DGICYT under grant number 
PB93-0048.

\clearpage

\figcaption[letfig.eps]{Left panel shows the integrated intensity map of
the HCO$^+$ J=1$\rightarrow$0 line carried out with the 30-m telescope
toward NGC 7023 (solid contours) superposed to the HI column density
image obtaining after combining VLA and DRAO data (grey scale)
(Fuente et al. 1993, Fuente et al. 1996). 
HCO$^+$ contours are 0.8 to 7.2 by 0.8 K kms$^{-1}$. 
The numbers in the wedge are in units of 10$^{20}$ cm$^{-2}$.
The star indicates the position of HD 200775, the white triangle indicates
the PDR peak and the filled black square, the molecular peak. The HCO$^+$
molecular filaments as observed with the IRAM PdB interferometer
are the thick contours, black is the filament at 2.4 kms$^{-1}$
and white the one at 4 kms$^{-1}$ (Fuente et al. 1996).
On the right we show (from top to down) the spectra of 
the HCO$^+$ J=1$\rightarrow$0, H$^{13}$CO$^+$ J=1$\rightarrow$0
CO$^+$ 
N=2$\rightarrow$1 J=5/2$\rightarrow$3/2 and 
J=3/2$\rightarrow$1/2 lines toward the molecular peak and 
the PDR peak.
Several channels of the original CO$^+$ spectra have been averaged.
\label{plate L1}}

\begin{table*}
\begin{center}
\caption{Observational parameters \label{tbl-1}}
\begin{tabular}{llrrrrr}
\multicolumn{1}{c}{Position} & 
\multicolumn{1}{c}{Molecule} & 
\multicolumn{1}{c}{Frequency} & 
\multicolumn{1}{c}{$\int T_{MB} dv$} & 
\multicolumn{1}{c}{v$_{lsr}$} & 
\multicolumn{1}{c}{$\Delta v$}  & 
\multicolumn{1}{c}{T$_{MB}$} \\ 
 & & \multicolumn{1}{c}{(MHz)} & 
\multicolumn{1}{c}{(K km s$^{-1}$)} & 
\multicolumn{1}{c}{(km s$^{-1}$)} & 
\multicolumn{1}{c}{(km s$^{-1}$)} &
\multicolumn{1}{c}{(K)} \\ 
\tableline
PDR peak & CO$^+$ & 235789.64 & 0.17(0.03) & 3.4(0.2) & 2.1(0.4) & 0.076  \\
R.A.:21$^h$ 01$^m$ 32$^s$.6    & 
CO$^+$ & 236062.55 & 0.23(0.02) & 2.7(0.1) & 2.1(0.4) & 0.104  \\
Dec: 68$^\circ$ 10' 27'' 
& CH$_3$OH & 239746.25 & $<$0.2\tablenotemark{a} & & & \\
(2000)   & CS     &  97980.97 & 0.92(0.04) & 2.3(0.1) & 0.7(0.1) & 1.31 \\
         & CS\tablenotemark{b} & 146969.05 & 1.38(0.02) & 2.3(0.1) & 0.7(0.1) 
 & 1.85 \\
         & CS\tablenotemark{c} & 244935.61 & 0.79(0.03) & 2.3(0.1) & 0.5(0.1) 
 & 2.73 \\
         & H$^{13}$CO$^+$     & 86754.33 & 0.34(0.04) & 2.38(0.04) &
0.5(0.1) & 0.61  \\
Mol. peak & CO$^+$ & 235789.64 & $<$0.03\tablenotemark{d} & & & \\
R.A.:21$^h$ 01$^m$ 31$^s$.6
& CO$^+$ & 236062.55 & $<$0.03\tablenotemark{d} & & & \\
Dec: 68$^\circ$ 11' 12''
 & CS     &  97980.97 & 1.02(0.02) & 2.6(0.1) & 0.7(0.1) & 1.36 \\
(2000)    & CS\tablenotemark{b} & 146969.05 & 0.64(0.02) & 2.7(0.1) & 0.7(0.1) 
 & 0.9 \\
         & CS  & 244935.61 & $<$0.04\tablenotemark{d}&&& \\
         & H$^{13}$CO$^+$      & 86754.33  & 1.10(0.01) & 2.8(0.1) &
0.7(0.1) & 1.46  \\
         & H$^{13}$CO$^+$      & 260255.48 & 0.44(0.01) & 3.0(0.1) &
0.4(0.1) & 1.04  \\
\end{tabular}
\end{center}
\tablenotetext{a}{3$\sigma$ upper limit assuming a linewidth of 2 kms$^{-1}$}
\tablenotetext{b}{Spectrum after degrading the angular resolution of
the CS 3$\rightarrow$2 map to have
that of the CS 2$\rightarrow$1 map.}
\tablenotetext{c}{The same as b but for the CS 5$\rightarrow$4 map}
\tablenotetext{d}{3 $\sigma$ upper limit assuming a linewidth of 1 kms$^{-1}$}

\end{table*}

\clearpage
\begin{table*}
\begin{center}
\caption{Integrated intensities, column densitites and abundance
estimates toward the PDR peak per velocity interval \label{tbl-2}}
\begin{tabular}{lccc}
\multicolumn{1}{c}{ } &  
\multicolumn{1}{c}{ 0 - 1.6 kms$^{-1}$} &
\multicolumn{1}{c}{ 1.6 - 3.2 kms$^{-1}$} &
\multicolumn{1}{c}{ 3.2 - 6.0 kms$^{-1}$} \\ 
\tableline
I(HCO$^+$ J=1$\rightarrow$0) (K kms$^{-1}$) & 
0.59(0.02)\tablenotemark{a} & 4.67(0.02) & 1.10(0.03)  \\
I(H$^{13}$CO$^+$ J=1$\rightarrow$0) (K kms$^{-1}$) & 
0.11(0.04) & 0.34(0.04) & $<$0.15 \\
I(CO$^+$ N=2$\rightarrow$1 J=5/2$\rightarrow$3/2) (K kms$^{-1}$) & 
0.02(0.01) & 0.14(0.01) & 0.07(0.02)  \\
I(CO$^+$ N=2$\rightarrow$1 J=3/2$\rightarrow$1/2) (K kms$^{-1}$) & 
$<$0.06 & 0.08(0.02) & 0.09(0.03)  \\
$\frac{I(CO^+ N=2\rightarrow1 J=3/2\rightarrow1/2)}
{I(CO^+ N=2\rightarrow1 J=5/2\rightarrow3/2)}$ &
        & 0.6(0.2) & 1.3(0.8) \\
N(HCO$^+$) (cm$^{-2}$) & 5 10$^{11}$ & 4 10$^{12}$ & 9 10$^{11}$  \\
N(H$^{13}$CO$^+$) (cm$^{-2}$) & 8 10$^{10}$ & 3 10$^{11}$ &  \\

N(CO$^+$) (cm$^{-2}$) & 3 10$^{10}$ & 2 10$^{11}$ & 1 10$^{11}$ \\
$\frac{N(CO^+)}{N(HCO^+)}$ & 0.01\tablenotemark{b} & 0.02\tablenotemark{b}
 & 0.11 \\
X(CO$^+$)\tablenotemark{c} & 4 10$^{-12}$ & 8 10$^{-12}$ & 4 10$^{-11}$ \\
\tableline 
\end{tabular}
\end{center}
\tablenotetext{a}{The number in parenthesis is $\sigma$}
\tablenotetext{b}{In these cases, the HCO$^+$ column density 
has been estimated from 
H$^{13}$CO$^+$ data assuming an isotopic ratio of 40 }
\tablenotetext{c}{Assuming X(HCO$^+$) $\sim$ 4 10$^{-10}$}
\end{table*}

\clearpage


%
%
\end{document}